# Comparing value of travel time and value of travel time saving with heterogeneity in travelers

Lijun Yu[1], and Baojun He[1]

[1]*School of Civil Engineering and Transportation, South China University of Technology, Guangzhou 510641, China*

E-mail: cthebaojun@mail.scut.edu.cn

## Abstract

In research on the value of past time, the value of travel time and the value of saving travel time are two different concepts that have been vaguely distinguished over an extended period of time. This paper applies the theory of perspectives to discuss differences in the value of travel time and savings in travel time between different respondent groups and under different models. To this end, this paper designs an RP-SP questionnaire for urban travel behaviour, and collects data on the travel preferences of 409 Guangzhou residents. By introducing potential profiling to capture the effects of heterogeneity between classes, the respondent group is divided into three categories according to its individual socio-economic characteristics such as age, income, educational attainment, etc. After that, the utility function of travelers is established, and the MNL model, MIXL model, S-MNL model, G-MNL model, LCL model and MM-MNL model are selected for the total sample and three types of groups to estimate the time value. This paper emphasizes the definitions of VTT and VTTS, and presents a method for estimating VTT and VTTS using LPA to identify traveler heterogeneity.

**Keywords:** travel time value, travel time saving value, prospect theory, latent profile analysis, MNL model, MIXL model, S-MNL model, G-MNL model, LCL model, MM-MNL model

# 1 Introduction

The value of travel time (VTT) is a concept extended by the value of time (VOT) in the domain of transport economy. VTT is important because it is a key element of cost analysis in most transportation project investments, and is also an important parameter in transportation planning models that influence transportation policymaking. In addition, from the perspective of social welfare, the benefits of VTT are huge. For example, the time value benefits of highway projects in China can account for 25%-50% of monetary benefits, while in developed countries, up to 80% of the benefits of transportation projects come from time value benefit.

However, VTT studies are controversial, in part due to a number of conceptual uncertainties. Value of time saving (VTTS) and VTT are also highly valued terms, but the difference between the two is often overlooked. Through the collection and comparison of existing literature, it is found that researchers seem to be more inclined to use VTT or VTTS (or VOT) alone in most studies. Why? To this end, our paper hopes to provide an answer to distinguish the difference between VTT and VTTS through the following three steps. First, design the RP-SP questionnaire and collect the data on the travel preference of residents in Guangzhou for preliminary analysis; second, use potential profile analysis to divide the types of respondents; third, based on prospect theory, introduce WTP and For WTA, a variety of Logit models were used to estimate and compare VTT and VTTS.

# 2 Literature Review

The earliest concept of VTTS comes from the Value of Time Saving (VTS) proposed by De Serpa[1]. Inheriting the idea of utility in time distribution theory, De Serpa creatively establishes a utility maximization model constrained by the minimum time required to consume commodities, and divides time value into three categories: Value of Time as Resource, commodity Value of Time as Commodity and Value of Time Saving (VTS). Among them, VTS is defined as the maximum amount that people are willing to pay to reallocate time between two alternative activities, so mathematically VTS is equal to the difference between VTR and VTC. Accordingly, in the transport economy, Jara-Díaz[2] similarly states that VTTS is equal to the Value of Leisure (VOL) minus the Value of Time Assigned to Travel (VTAT). Later, scholars such as Hössinger[3] and Tang[4] found that the size of VOL mainly depends on the personal income of the traveler, while estimating the VTAT also needs to comprehensively consider the traveler's socioeconomic attributes, travel mode, travel purpose and traffic service quality. And other factors. In addition, how to use time during travel also affects VTAT[5]. For example, people who can effectively use travel time, such as mobile office during travel, have significantly lower VTTS[6]. Recently, Basil[7]collected questionnaire data from Zurich workers and obtained VTAT based on estimated VTTS and VOL. Their results suggest that more emphasis should be placed on travel comfort in transportation investments.

Therefore, in the above study, VTTS can be defined [8] as the monetary value of the part of the time that the traveler chooses faster travel mode brings about the reduction. Therefore, Jara-Díaz[2, 8] and Mackie[9]argue that the realization of VTTS should also depend on what activities the saved time is used for. Simply put, assuming that the marginal utility of market working time is equal to 0, if this part of the time saved is invested in productive economic activities, then VTTS can be regarded as a generalized wage income, while if it is used for leisure and entertainment, then VTTS may be biased. Low is even equal to 0. Correspondingly, VTT emphasizes the currency converted by the traveler's time spent in travel, which is equivalent to VTAT. It can be seen that the VTT and VTTS here reflect an opportunity cost linked to the traveler's income and travel purpose. However, since VTT (or VTAT) has the disadvantage that it can only be estimated indirectly through VOL and VTTS, and VTTS seems to be a more direct measure of the time benefit between different modes of transportation and between new and old traffic construction Therefore, recent studies prefer to use the term VTTS instead of VTT, especially in transportation policy documents. The Florida Department of Transportation Research Center[10] collected trip-specific data on Express Lane 95 in Miami, Florida from 2010 to 2011 and found that these travelers had an estimated VTTS of about 49% of their annual household income and hourly earnings, the report said. Contains information on several existing projects across the United States; Wardman et al.[11] were commissioned by the British Department for Transport to write a report on the time-saving value of business travel in 2013; the University of Leeds and the British Department for Transport[12] in A final report of the study "Providing Market Research for Time Saving and Reliability Value" was completed in 2015.

Small[13] and Huq[14] and others believe that the definition of VOT in transportation economics is actually open, which depends on the content of the study, so there are many different measures of VOT, which is consistent with our point of view . Another mainstream interpretation of VTTS is the willingness-to-pay (WTP) of travelers for the unit travel time saved. At present, there are also many studies supporting the interpretation of VTTS into WTP, such as Bliemer[15], Flügel[16] and Merkert[17]. We believe that VTTS under this definition means that "time" is a tradable "commodity", but this "commodity" can only be gained and lost through the comparison of different travel modes or different travel conditions. Specifically, there are two modes of transportation (such as bus and private car), one is faster but more expensive, the other is slower but cheaper, if the traveler is willing to choose the faster mode of transportation, he will choose the comparison Slow transportation saves some time, and extra money pays for the "time" saved, so VTTS can be simply expressed as . Of course, the actual travel choice behavior is very complex, and the heterogeneity of travelers and changes in travel conditions directly affect the size of VTTS. We have the following conclusions: The two definitions of VTTS do not negate each other. The first VTTS emphasizes the value of the time saved in productive activities or leisure activities, and the second VTTS emphasizes savings by direct monetary payment. time.

Further, as an economic term paired with WTP, Willingness-to-Accept (WTA) refers to the minimum amount of money that consumers need to compensate for the loss of goods they hold. In standard economic theory, WTP and WTA are the same amount of money for an object[18], leading to some unnecessary ambiguity about VTTS and VTT. Shao[19] uses the directly available WTA in the form of difference (as in the previous example of WTP) to analyze VTTS, and the WTA he uses is not an estimate in the true sense. In fact, a large number of social experiments show that WTP and WTA are always asymmetric and have , and the first derivative of WTA is WTP[20]. In short, individuals are more averse to losses than they like to gain in the face of their current holdings, a phenomenon also known as the "endowment effect". A long time ago, behavioral economists represented by Kahneman[21] and Tversky[22] noticed this phenomenon and believed that the basis of individual value evaluation was not the absolute value of the possible decision-making, but a certain existing value. The psychological neutral reference point is used as the standard, and the evaluation result is understood as the degree of deviation between the actual situation and the reference point. Bateman et al. [23] designed an experiment to test the effect of reference dependence, and the results of this experiment fully support the famous prospect theory and its extended model proposed by Kahneman et al. [21]. Prospect theory and the "endowment effect" reflect the general human psychological state of "loss aversion" and "risk aversion", which can be seen everywhere in human social activities. Knetsch et al.[24] distributed candies and cups among the investigated student groups and allowed them to exchange with each other, and found that most students refused to exchange the original items; Hardie et al.[25] found that consumers usually like to put the goods The price or quality is compared with the price or quality of the same product that was last consumed, and the negative utility brought by the price increase or quality decrease exceeds the positive utility brought by the price decrease or quality increase; Shogren et al.[26] designed three different It is found that there is also an "endowment effect" in the initial bidding of the auction, but this effect may be eliminated as the number of bidding increases. On the other hand, there are also many literatures that further contribute to the conclusion of the asymmetry of WTA and WTP. Zhao et al.[27] introduced uncertainty, irreversibility, and limited cognitive opportunity commitment cost to explain WTP and WTA; Horowitz et al.[28] estimated that the ratio of WTA/WTP was too high, implying that the income effect is very high high, inconsistent with neoclassical preferences.

Similarly, prospect theory and the "endowment effect" are also applicable to the analysis of traffic travel behavior. The reference point can be the travel mode of the same period, the same vehicle, old or new, or the travel route. The reference point is usually an expectation and decision-making basis from the traveler's heart, and is generally closely related to the traveler's own travel experience, travel environment, and the level of transportation technology development. A traveller's VOT may not be the absolute value of the utility due to travel time and expense, but rather a utility bias based on a neutral reference point that already exists in their brains. At the same time, it should be noted that for a traveler, the loss or gain of the travel time value is judged according to the reference travel mode. When the reference point changes, the loss or gain of its VOT will also follow. change. De Borger & Fosgerau[29, 30] verified this view mathematically. He introduced the value function of travel time and cost without reference point, combined with the arguments of Bateman et al.[23], and finally obtained the relationship between WTA, WTP, EL and EG. On the basis of the theory proposed by Borger, Ramjerdi & Dillén et al.[31] conducted MNL model and RPL model estimation on the SP data collected from travel in Sweden, and obtained the conclusion that the value of WTA is 1.5~2 times the value of WTP, which is The first practical study to estimate VOT using WTP and WTA. Hess et al.[32] examined the traveler's personal socioeconomic attributes and travel attributes can lead to the asymmetry of choice preferences, that is, there is a difference between WTA and WTP. Subsequently, Hess et al.[33] proposed a heterogeneity and reference point-dependent travel time value research framework. Hess & Daly[34] then discussed the relationship between VTTS and VTT, and believed that most of the current research results misused VTTS. They give the result that VTTS is equivalent to Value of gain, and VTT is the result of the geometric mean of Value of gain and Value of loss. Liu et al.[35] analyzed the calculation principle of traveler's utility from the perspective of psychological accounting, and proposed a travel choice behavior model (MA-TC model) considering travel time and cost based on prospect theory.

So far, we can clearly recognize that the premise of correctly distinguishing VTTS and VTT is to recognize the asymmetry of WTP and WTA, which is one of the research work on VOT by Hess, Daly, Borger and others in recent years. To sum up, there are still few differential works specifically targeting VTTS and VTT. On the other hand, VOT studies are very concerned with the heterogeneity of groups including income, age, education level, etc., but not many have applied latent class/profile analysis to reveal the heterogeneity.

# 3 Theoretical framework

## 3.1 G-MNL model, LCL model and MM-MNL model

The Discrete Choice Model (DCM) has been the primary mathematical tool for scholars to assess the value of time for the past few decades. McFadden[36] first established a complete multinomial Logit (Multinomial Logit, MNL) model in 1974. The model is simple and practical, but there is the limitation of IIA (Independence of Irrelevant Alternatives), which means that the preferences of different individuals are the same. In order to solve the problem of IIA limitation, the multinomial Probit (Multinomial Probit, MNP) model[37] and the mixed polynomial Logit model (Mixed Logit, MIXL), also known as the random parameter Logit model (Random Parameter Logit, RPL) model[38] The heterogeneity of individual preferences is explained by setting random parameters that obey a certain distribution. Train[39] summarizes the MNL and MIXL models.

Most discrete-choice models, including MNL and RPL, are applied assuming that the weights of the parameters of each utility relative to the program attributes follow a multivariate normal distribution (VMN) in the population. However, Louviere et al.[40] considered MVN unreasonable, because they observed that the selection behavior of some individuals was more random, and the parameter weights of all their attributes would be enlarged or reduced in order, which is a scale effect. To this end, Fiebig et al.[41] proposed a Generalized Multinomial Logit (G-MNL) model for revealing parameter scale and heterogeneity.

In the discrete choice model, it is assumed that the individual $n(n=1,2,...,N)$ chooses alternative $i(i=1,2,...,I)$, and the individual $n$ utility function is composed of observable utility $V_{nk,i} = \beta x_{nk,i}$ and error term $\varepsilon_{nk,i} \sim$ i.i.d, which is expressed as

$$U_{nk,i} = V_{nk,i} + \varepsilon_{nk,i} = \beta x_{nk,i} + \varepsilon_{nk,i} \tag{1}$$

where, $x_{nk,i}$ is the observable attribute in the alternative scheme and its parameter $\beta$. Fiebig et al.[41] gave the form of parameter correction $\beta_n$ in G-MNL:

$$\beta_n = \sigma_n \beta + \gamma \eta_n + (1-\gamma)\sigma_n \eta_n \tag{2}$$

Therefore, the utility function of G-MNL model is expressed as:

$$U_{ni} = [\sigma_n \beta + \gamma \eta_n + (1-\gamma)\sigma_n \eta_n]x_{ni} + \varepsilon_{ni} \tag{3}$$

where, $\sigma_n$ is the scale error term (weight) of the parameter; $\eta_n$ is the deviation average of the parameter; $\gamma$ controls how the variance of the remaining heterogeneity varies with size. If the parameters in $\beta_n$ change, we will get two forms of g-mnl model:

• If $\gamma = 1$, $\beta_n = \sigma_n \beta + \eta_n$ and it is G-MNL-I model;

• If $\gamma = 0$, $\beta_n = \sigma_n(\beta + \eta_n)$ and it is G-MNL-II model;

Based on the above two forms of G-MNL model, S-MNL model, MIXL model and original MNL model can be obtained by constraining the parameters $\beta_n$:

• There is G-MNL-I model, if $\mathrm{var}(\eta_n) = 0$, $\beta_n = \sigma_n \beta$ and it is S-MNL model;

• There is G-MNL-II model, if $\sigma_n = 1$, $\beta_n = \beta + \eta_n$ and it is MIXL model;

• Especially, if $\mathrm{var}(\eta_n) = 0$ and $\sigma_n = 1$, $\beta_n = \beta$ and it is orignal MNL model.

Similar to the MIXL model, the parameters of the G-MNL model do not have closed-form solutions, and simulation methods are usually used to solve them. The Monte Carlo method based on the law of large numbers and the central limit theorem is a common method for resolving such patterns. According to Greene's work [42], we can give the expression for the log-likelihood function of the data samples:

$$\log L = \sum_{n=1}^{N} \log \left\{ \frac{1}{R} \sum_{r=1}^{R} \prod_{k=1}^{K} \prod_{i=1}^{I} P(i, x_{nk}, \beta) \right\}^{\delta_{nk,i}} \tag{4}$$

where, $R(r=1,2,...,R)$ is the number of simulations; If the individual chooses the alternative $i$, $\delta_{nk,i}=1$, otherwise; $P(i,x_{nk},\beta)$ is the probability of individual $n$ selecting the alternative $i$ in the case $k$, and the expression is:

$$P(i,x_{nk},\beta)=\frac{\exp\left[\sigma_d\beta+\gamma\eta_d+(1-\gamma)\sigma_d\eta_d\right]x_{nk,i}}{\sum_{i=1}^{I}\exp\left[\sigma_d\beta+\gamma\eta_d+(1-\gamma)\sigma_d\eta_d\right]x_{nk,i}} \quad (5)$$

where, $\sigma_d=\exp(\bar{\sigma}+\tau\varepsilon_d)$ in which $\varepsilon_d\in N(0,1)$; $\eta_d\in \text{MVN}(0,\Sigma)$. Both $\sigma_d$ and $\eta_d$ are parameters estimated by simulation。

In addition to the G-MNL model, another model has become popular in recent years, the Latent Class Logit (LCL) is an extension of the MNL model for classifying data for heterogeneity. Absorbing the advantages of the LCL and MIXL models, Keane[43], Hensher[44] and others introduced the MM-MNL model. By restricting the conditions, the model can degenerate into a LCL or a MIXL model. In the same way, the expression of the log-likelihood function of the data samples in the MM-MNL model is given:

$$\log L=\sum_{n=1}^{N}\log\left\{\sum_{c=1}^{C}\pi_q(\theta)\frac{1}{R}\sum_{r=1}^{R}\prod_{k=1}^{K}P\left[\delta_{nk,i}\left|(\beta+w_{n,k}),x_{nk}\right.\right]\right\} \quad (6)$$

where, the classification probability is $\pi(\theta)_q=\dfrac{\exp(\theta_q)}{\sum_{q=1}^{Q}\exp(\theta_q)}$ ($q=1,2,...Q$; $\theta_Q=0$);

The meaning of other mathematical symbols is the same as the log likelihood function of G-MNL model. The probability of individual $n$ selecting the alternative $i$ in the case $k$ is:

$$P\left[\delta_{nk,i}\left|(\beta+w_{n,k}),x_{nk}\right.\right]=\frac{\exp\left\{\sum_{i=1}^{I}\exp\left[\delta_{nk,i}(\beta+w_n)\right]x_{nk,i}\right\}}{\sum_{i=1}^{I}\exp\left\{\sum_{i=1}^{I}\exp\left[\delta_{nk,i}(\beta+w_n)\right]x_{nk,i}\right\}} \quad (7)$$

where, $w$ is uncorrelation coefficient of data samples.

## 3.2 Formulation of value of time

For the analysis of value of time, the utility function of traveler's choice of travel mode is mainly composed of travel mode attribute items (time, cost, reliability, etc.) and traveler attributes (gender, age, income, etc.). Therefore, under the discrete choice model, the travel time value of the travelers in the sample is expressed as:

$$\text{VOT} = \frac{\partial U / \partial Cost}{\partial U / \partial Time} \tag{8}$$

The above formulation of VOT is the same as that of VTTS, but the data samples are different. For details, see the data analysis section.

## 3.3 Latent profile analysis

Latent Class/Profile Analysis (LCA/LPA) is an emerging individual-centered statistical analysis method, which uses latent class variables to explain the association between external continuous variables and realizes the relationship between explicit variables. local independence. Compared with the LCL model and the MM-MNL model, LCA/LPA has great advantages, which will be explained in detail in the data analysis section. This method is widely used in humanities, social sciences and medicine. In the analysis of traffic travel behavior, Greene & Hensher[45] used this method earlier to divide travelers with similar characteristics into several independent groups and then use MXL Models study their choice behavior for long-distance car travel on three road types. Recently, many scholars have used this method to explore more issues in travel behavior. For example, Molin et al.[46] used LCA to identify the attitudes of public transport or bicycle drivers towards public transport, while Rafiq et al.[47] Furthermore, the survey groups were divided by race, age, and travel purpose, and transportation-based activity-travel patterns were analyzed. Such research can help governments and operating agencies to formulate sustainable transportation policies; Haas et al.[48] has an new analysis of potential transitions is applied within the resident life record framework to assess how life events influence travel behavior.

LPA is an extension of LCA on continuous explicit variables. The principles and steps are the same as those of traditional LCA. The difference is that LPA does not require dummy variable data, and the probability distribution is expanded into a density distribution in the expression.

In LPA, the variance of continuous index is decomposed into the variance between sections and within sections, so the variance of the index $i$ in section $k$ can be expressed as:

$$\sigma_i^2 = \sum_{k=1}^{K} P(c_i = k)(\mu_{ik} - \mu_i)^2 + \sum_{k=1}^{K} P(c_i = k)\sigma_{ik}^2 \qquad (9)$$

where, $\mu_{ik}$, $\sigma_{ik}^2$ is mean and variance of index $i$ in section $k$; $P(c_i = k)$ is the probability of the section $k$, that is, the proportion of individuals in the section to the whole.

Local independence and homogeneity are two basic assumptions of LPA / LCA model. Local independence means that any two observation indexes in the profile are not related, that is $P(c_i = k) = \prod_{i=1}^{I} P(y_i | c_i = k)$. When the hypothesis of and homogeneity is satisfied, the variance in the profile of formula (9) is 0, and the probability density function (LPA model) of the significant variable vector $y_i$ is expressed as

$$f(y_i) = \sum_{k=1}^{K} p(c_i = k) f(y_i | c_i = k) \qquad (10)$$

The core function of LPA/LCA is to determine the number of categories. This paper uses the four most commonly used information indicators as the basis for the number of profiles, namely: AIC (Akaike Information Criterion), BIC (Bayesian Information Criterion), Entropy and BLRT (Bootstrapped) Likelihood Ratio Test).

## 4 Questionnaire Design and Data Collection

The study of the transport preferences of Guangzhou residents in the city is conducive to the development of time value research. As a city with a developed economy and a large population, Guangzhou's advanced urban transportation system has the characteristics of huge demand for residents' travel, high dependence on public transportation, and various travel modes. According to reports[49], in recent years, the total number of motorized trips in Guangzhou has exceeded 20 million per day, while the average daily passenger flow of Guangzhou's public transport in 2020 will be 11.56 million, of which the subway accounts for 57%. At the same time, it is expected to increase the proportion of public transportation in the city to more than 60% of motorized travel by 2035. Second, there are different modes of transportation in the city, and shared bikes, online calls, and electric vehicles have become important modes of transportation for residents. In addition, the Guangzhou Municipal Government's website contains detailed information on public traffic, population and the economy, as well as related information. As a result, we can collect sufficient reliable and diverse samples within a short period of time and conduct a comparative analysis.

The questionnaire of this paper adopts the combination of RP survey and SP survey, and is divided into three parts: the first part is the traveler's personal socioeconomic characteristics information, the second part is the traveler's most recent travel characteristics information, the third Part of it is the choice of travel mode in virtual scenarios. Here, since the time value of walking is difficult to measure, and each mode of transportation must be accompanied by walking, the RP-SP survey does not consider walking as an independent mode of travel.

This paper uses "Wenjuanxing" to make questionnaires, and distributes the questionnaires to travel groups of different ages in Guangzhou through Internet social media and face-to-face. In order to test whether the questionnaire can adapt to the data requirements and take care of the filling experience of the respondents, a three-day pre-survey will be conducted from September 25 to September 27, 2021, and 69 samples will be collected. According to the preliminary investigation, the content of the questionnaire will be adjusted appropriately, and the formal investigation will be carried out from October 3 to October 7, 2021. The total number of valid questionnaires collected was 409. Figure 1 shows the process by which respondents respond to the questionnaire.

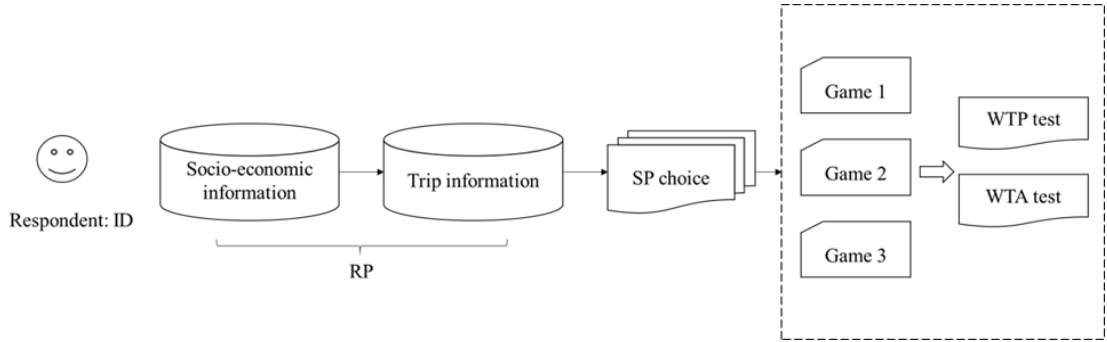

Figure 1. Data collection process

Personal socioeconomic information and travel information are the two components of the questionnaire RP survey. According to the RP survey design and pre-survey results of the questionnaire in the existing research, the first part of our questionnaire[50-55] includes gender, age, personal (family) income, education level, occupation, driver's license, and car ownership; The second part[50-58] includes the purpose of travel, frequency of trains, number of people traveling, travel time (in-vehicle time and out-of-vehicle time), and travel cost. In recent years, with the concept of Mobility as a Service (MaaS) proposed, the quality of services such as the comfort, safety, convenience, reliability, flexibility, and riding environment of transportation has received much attention[52-54, 57, 58], which mainly reflects the traveler's subjective feelings about travel, so we add a five-level Likert satisfaction score question in the second part. In other details, since some travelers are not the main body of travel mode who need to pay related expenses, for example, residents over 65 years old in Guangzhou can take the bus for free, so the option of "free/no payment" has been added to the cost topic to enable correspond to the actual situation. In addition, the cost of self-driving and carpooling in real life will be significantly higher than other modes of travel, so respondents who choose these two modes of travel will jump to a question with a higher cost to answer.

Table 1 and Figure 2 present the preliminary statistical analysis of the samples for the first two parts. The results show that the overall respondents are more middle-aged and young (71.15% of the total 23-45 years old), have a higher education level, and have a decent and stable job. For the most recent trip, respondents favored low-carbon travel methods, while also relying heavily on public transportation, which is in line with the current travel situation in China. It is worth noting that more than 30% of households own bicycles (electric vehicles), and bicycles (electric vehicles) account for 20.05% of trips and have a higher proportion in short- and medium-distance travel. It shows that flexible, convenient and cheap bicycles, particularly electric vehicles, are one of the main means of transport for short and medium distance trips in the city.

Table 1. Summary statistics of socio-economic characteristics

| Characteristic | Description | Code | Sample |
|---|---|---|---|
| Gender | Male | 1 | 50.61% |
|  | Female | 0 | 49.39% |
| Age | 12-22 | 1 | 17.36% |
|  | 23-35 | 2 | 32.76% |
|  | 36-45 | 3 | 26.65% |
|  | 46-55 | 4 | 11.74% |
|  | 56-65 | 5 | 6.85% |
|  | >65 | 6 | 4.65% |
| Income | No fixed income/No income | 1 | 17.85% |
|  | 1500-3000 yuan | 2 | 17.60% |
|  | 3000-5000 yuan | 3 | 19.32% |
|  | 5000-10000 yuan | 4 | 28.85% |
|  | 10000 yuan | 5 | 16.38% |
| Education | Below high school | 1 | 12.47% |
|  | High School | 2 | 23.96% |
|  | Bachelor | 3 | 45.23% |
|  | Master | 4 | 14.18% |
|  | PhD | 5 | 4.16% |
| Job | Fixed employment | 1 | 31.05% |
|  | No fixed employment/ Self-employment | 2 | 24.69% |
|  | Unemployment/Retired | 3 | 19.32% |
|  | Student | 4 | 24.94% |
| Car Ownership | None | 1 | 10.51% |
|  | Bike or e-bike | 2 | 34.23% |
|  | One car | 3 | 46.45% |
|  | More than one car | 4 | 8.80% |
| Population of family | 2 and below | 1 | 6.85% |
|  | 3 | 2 | 57.70% |
|  | More than 3 | 3 | 35.45% |
| Car license | Yes | 1 | 46.70% |
|  | No | 0 | 53.30% |

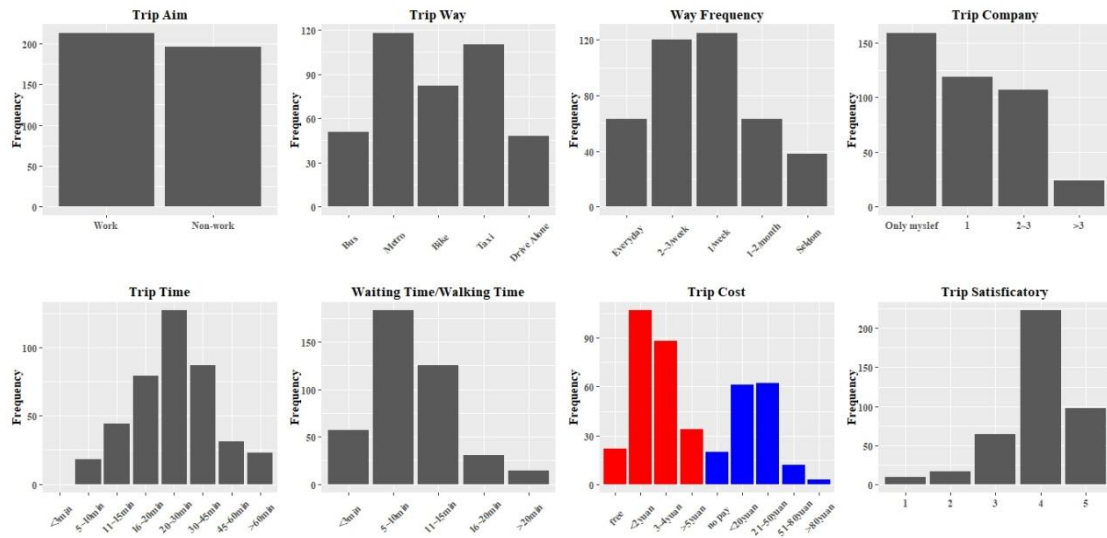

Figure 2. Summary charts of travel characteristic

*Figure 2. Trip Cost: Red bars stands for frequency of cost by bus, metro and bike, blue bars stands for other two modes.

The SP survey is a method of measuring respondents' preference for alternatives based on virtual scenario choices. Before proceeding with the SP survey design, two points need to be noted. First, a number of studies have shown that the difference in the respondents' subjective awareness of treating virtual situations and real situations will lead to a large deviation in the estimated time value. The VOT estimated by Small et al.[59] and Isacsson et al.[60] RP survey is higher. Therefore, in order to reduce this deviation, the factors of the assumed travel mode (time, cost, reliability, etc.) in the SP survey design should match the actual situation as much as possible with each travel mode. , which is consistent with Li[61]. Secondly, the information involved in the design of the SP survey should not be too much, otherwise it will affect the respondents' judgment on the preferred travel mode, especially for the elderly[62]. To this end, our SP survey designs three virtual scenarios and five alternative travel modes corresponding to each virtual scenario. Of these, Scenario 1 is a short-distance trip to the city, Scenario 2 is a mid-way trip to the city, and Scenario 3 is a long-distance trip to the city. All five modes of travel include travel time, waiting time, reliability and cost. Based on the most recent trip distance, respondents select the appropriate situation to respond. Table 2 shows the selection pool and attributes of the SP survey design.

Table 2. SP choice set

| Mode | Time[min] | Waiting time[min] | Reliability[min] | | Cost[yuna] |
|---|---|---|---|---|---|
| | | | Early | Late | |
| Bus | 25,50,100 | 5 | 3,5,10 | 10,15,25 | 2,2,4 |
| Metro | 18,35,70 | 3 | 3,5,8 | 3,8,10 | 2,4,7 |
| Bike(or E-bike) | 40,75,120 | 0 | 5,5,5 | 10,15,20 | 1.5,1.5,3 |
| Taxi(or carpool) | 10,25,45 | 2 | 3,5,15 | 5,10,15 | 15,25,50 |
| Drive alone | 10,25,45 | 0 | 3,5,15 | 5,10,15 | 8,15,30 |

After the respondents complete the choice of travel mode in the SP survey, two questions will be asked to understand the respondents' willingness to pay and willingness to pay for travel modes. Specifically, taking the time and cost of the selected row approach as reference points, question 1 asked respondents to select two values between the 5%-80% level as the degree of time savings and overpayment; question 2 Respondents were asked to choose two values between the 5%-80% level as the degree of spending more time and paying less. The distribution of the samples is shown in Figure 3.

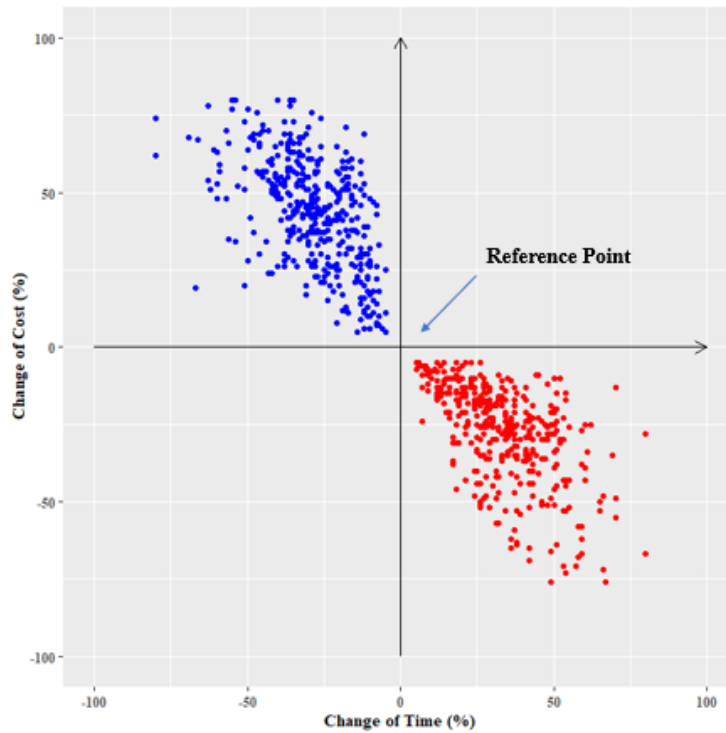

Figure 3. Scatter diagram of WTP&WTA test

# 5 Data analysis and results

## 5.1 Latent Profile Analysis of Respondents

Here are the four indicators AIC, BIC, Entropy and BLRT used to evaluate the goodness of fit of different LPA profiles. The indicators generally have the following criteria[63]: The smaller the AIC and BIC values, the better the model fit is. ; The larger the Entropy value, the better the model fit and the more sensitive[64]. The Entropy value is between 0 and 1. When Entropy is used, more than 20% of the individuals have classification errors. Entropy indicates that the classification accuracy rate exceeds 90%. . Simulation studies such as Nylund[65] found that for small samples, BLRT has sufficient test power. Compared with LMR and BIC, BLRT can reduce the possibility of type 1 errors and increase the accuracy of model recognition. When BLRT this metric is , the model representing the category is better than the model of the category.

In this paper, the "mclust"[66] package of R is used to perform LPA on the first part of the samples with codes collected from the questionnaire. The AIC and BIC of the three types are both smaller and the difference between BIC and the fourth type is not large, and the Entropy value is greater than that of the second and fourth types. In addition, Howard et al.[67] believe that if there is no obvious advantage of the profile in the classification index, then a parsimonious profile model should be selected. Therefore, considering comprehensively, it is better to divide it into three sections.

Table 3. Result of LPA for Part I

| Classes | AIC | BIC | Entropy | BLRT | Class probability(min/max) |
|---|---|---|---|---|---|
| 1 | 24376.55 | 24458.35 | 1.00 | \ | 1.00/1.00 |
| 2 | 23218.36 | 23346.17 | 0.93 | <0.01 | 0.96/0.99 |
| 3 | 22480.59 | 22654.41 | 0.96 | <0.01 | 0.97/0.99 |
| 4 | 21964.82 | 22184.65 | 0.95 | <0.01 | 0.93/0.99 |

After being divided into three categories, the mean scores of each category in the first part are plotted as a line graph, see Figure 4. According to the score of each category in each item, we named the type I population as "student group", accounting for 24.69% of the sample size, which is characterized by no stable income, but generally high education level; Type II population was named as "student group" Social backbone group", accounting for 43.52% of the sample size, which is characterized by a high and stable income occupation, high educational level, and generally have cars and driver's licenses; Type III group is named "older group", accounting for 31.79% of the sample size %, which is characterized by older age and whose main source of income is pensions. It should be noted that due to the question setting, we may collect up to 28,800 types of respondents, so there must be some respondents whose attributes are quite different from the attributes of the divided categories. For example, the Type I population may include some lower-income respondents. non-students. But on the whole, these three groups of people are representative and conform to the actual demographic characteristics of Guangzhou residents.

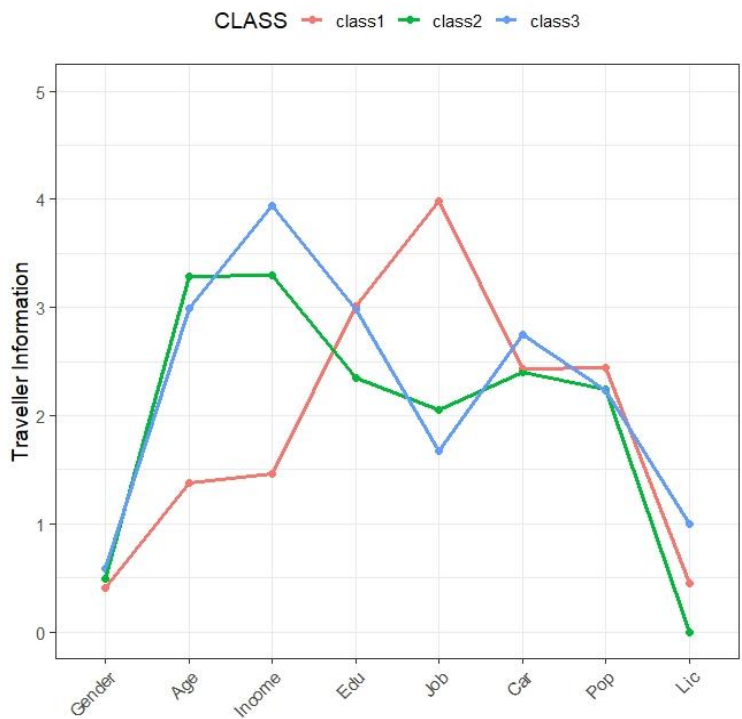

Figure 4. LPA result of three types

## 5.2 VOT

Before using the discrete choice model to estimate VOT, we can count some simple results related to VOT through sample data. A factor that is strongly linked to VOT is the traveller's personal income (or household income). There is a general belief that high-income individuals tend to choose faster, more comfortable, but also relatively expensive modes of transportation. It is worth mentioning that the importance of income is also reflected in: on the principle of opportunity cost, the direct conversion of a traveler's VOT into the wage rate during travel is one of the traditional VOT estimation methods. Therefore, although we use LPA to weaken the direct effect of income on VOT in the estimation, it is necessary to re-emphasize the analysis of income for each group.

In Chapter 4, we mentioned that each respondent took two willingness tests after choosing their own mode of travel under the SP scenario. In the first test, we identified each respondent's expected travel time savings and acceptable additional travel costs as planned travel savings for the current mode of transportation and additional travel time workable. Shao et al.[19] gave a non-model estimated WTA, which is expressed as the ratio of the cost difference between the two travel modes to the time difference, which is consistent with the definition of reference-free value of time in the inference of Borger's work[30]. In the same way, we also get a WTP and WTA result that does not need to be estimated using the model, and name it underlying WTP and underlying WTA:

$$\text{underlying } WTP = \frac{Cost_{\text{SP-loss}}}{Time_{\text{SP-gain}}} \tag{11}$$

$$\text{underlying } WTA = \frac{Cost_{\text{SP-gain}}}{Time_{\text{SP-loss}}} \tag{12}$$

It is important to note that underlying WTP and underlying WTA are calculated by comparing respondents' "ideal" and "actual" modes of travel in the SP survey, not between modes of travel.

The average monthly income of all respondents, type I group, type II group and type III group was 5351.83 yuan, 1187.62 yuan, 6256.74 yuan and 8555.83 yuan respectively. The overall monthly income of respondents was much higher than that of the Type I group and slightly lower than that of the Type II group. The National Bureau of Statistics announced that in 2021, the average weekly working time of employees in companies at the national level will be 46.3 hours. We calculate the underlying WTP and underlying WTA of this method separately from the total sample and the sample data of types I~III, and integrate the hourly wages of travelers. We get the box plot in CNY/h in Figure 5. The box plot The width is proportional to the sample size. The underlying WTA of Type III was 64.98 CNY/h, ranking first. The total sample population and the underlying WTP and underlying WTA of types I~III have many data outliers, and are much higher than the hourly wages of the sample, reflecting that some travelers have exchange perceptions of travel time and travel expenses and their income. Not necessarily consistent. On the other hand, it is interesting that the upper range of the underlying WTA is higher than both the hourly wage and the underlying WTP, and the underlying WTP tends to have a lower range lower than the other two.

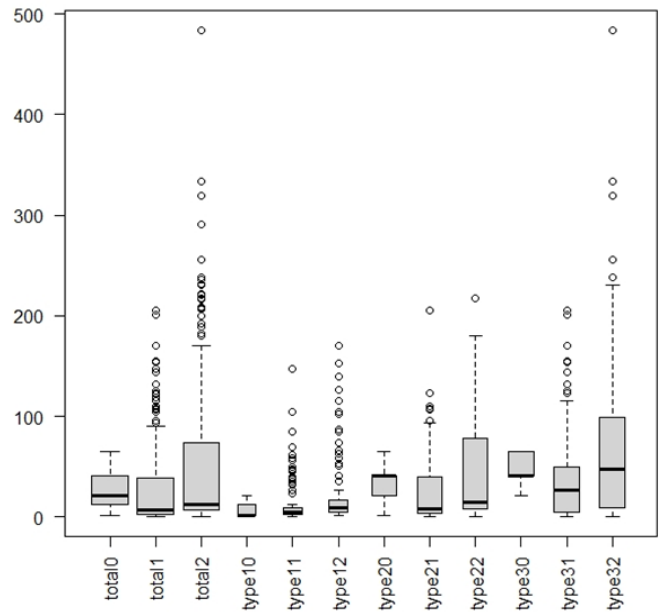

Figure 5. Income, underlying WTP and underlying WTA

This preliminary statistics can give us some intuitive feelings about the comparison of WTP and WTA with income, but the disadvantage is also obvious. It cannot well express the group's travel preferences and sensitivity to travel time and travel costs. Therefore, constructing a utility function and estimating VOT using a discrete choice model has been a common practice for many years, and it is also the main research tool of this paper, because it can reflect the willingness of the walker.

Below, we formally use a variety of Logit models to estimate model parameters. For this purpose, we use the "mlogit" command in STATA and the "CLOGIT", "RPLOGIT", "SMNLOGIT", "GMXLOGIT", "LCLOGIT" and "NLOGIT" commands at the same time. LCRPLOGIT" command. The purpose of using the two software is to compare whether the model results are consistent and stable. Use the help command to learn more about STATA's discrete choice model; for how to use NLOGIT, refer to the NLOGIT Reference Guide[68].

**5.2.1 Preliminary estimation of time value in RP survey**

STATA's "mlogit" command works on MNL models with unordered categorical dependent variables. The actual value of the dependent variable in the model is irrelevant. Therefore, this type of MNL model is also called a multiphase logistic regression model, which is essentially an extension of the binary Logit model. STATA's document[69] states that some people think that the MNL model should be a Conditional Logit model.

We use this type of MNL model to estimate the RP part of the questionnaire sample using public transport as the reference scheme. The results are shown in Table 4. According to statistics, the proportion of respondents who choose public transportation, subway, bicycle (electric vehicle), taxi and self-driving car is divided into 12.47%, 28.85%, 20.05%2, 26.89% and 11.74%. The results show that most of the coefficients estimated by this type of MNL model are not significant, mainly because the model contains too many terms and the relationship between the terms is weak, which makes it difficult to fit; or the data sample size is too small. Nonetheless, given a significance level of 10%, we can draw some solid conclusions: higher in-vehicle and out-of-vehicle time are likely to prompt respondents to choose public transport as a mode of travel, with higher costs Respondents are more inclined to choose the original mode of travel rather than public transport. In terms of travel frequency, the coefficient of taxis (1.245) is greater than that of bicycles (0.601) and self-driving cars (0.609), indicating that taxis are the least used means of transportation for respondents. In addition, respondents with stable incomes tend to choose comfortable modes of travel such as taxis and self-driving cars. Respondents with a driver's license and a car were significantly more likely to choose to travel by car. Of course, the biggest disadvantage of this type of MNL model is that the size and positive or negative of the coefficients may change under different reference schemes, but the general conclusion is still reliable. For example, when trying to use self-driving as a reference solution, it is found that the time and cost coefficients are both positive and negative compared to public transportation as a reference solution. However, in general self-driving takes less time and is more expensive, so we can still draw conclusions that are consistent with using public transport as the reference scenario. Here, the value of time inside the car is -4.67/-82.46/-7.09/-3.60 CNY/h, and the value of time outside the car is -5.43/-144.96/-18.38/-16.98 CNY/h.

Table 4. Result of estimating RP data with *MNL**

| Variables | Bus | Metro | Bike | Taxi | Drive Alone |
|---|---|---|---|---|---|
| | | | Coefficient | | |
| **Travel characteristic** | | | | | |
| In-vehicle time (min) | Base | -0.0239** | -0.0859*** | -0.0803*** | -0.0377* |
| | | (0.0114) | (0.0187) | (0.0222) | (0.0199) |
| Out-of-vehicle time (min) | | -0.0278 | -0.151*** | -0.215*** | -0.178*** |
| | | (0.0356) | (0.0493) | (0.0671) | (0.0663) |
| Travel cost (CNY) | | 0.307** | 0.0625 | 0.702*** | 0.629*** |
| | | (0.126) | (0.141) | (0.137) | (0.137) |
| Travel aim | | -0.283 | -0.374 | -0.352 | -0.554 |
| | | (0.310) | (0.341) | (0.414) | (0.428) |
| Travel frequency | | 0.145 | 0.601*** | 1.245*** | 0.609** |
| | | (0.206) | (0.232) | (0.298) | (0.290) |
| Travel companion | | 0.662** | 0.603** | 1.279*** | 1.203*** |
| | | (0.264) | (0.291) | (0.357) | (0.361) |
| Travel satisfactory (Likert) | | 0.0200 | 0.270 | 0.295 | 0.459 |
| | | (0.187) | (0.246) | (0.373) | (0.369) |
| **Personal characteristic** | | | | | |
| Gender | | 0.173 | 0.489 | -0.0297 | 0.0691 |
| | | (0.393) | (0.458) | (0.578) | (0.576) |
| Age | | -0.0162 | -0.0212 | 0.000923 | 0.00825 |
| | | (0.0135) | (0.0168) | (0.0220) | (0.0217) |
| Income(CNY) | | 0.000137 | 1.56e-05 | 0.000270** | 0.000168 |

|             | (9.50e-05) | (0.000105) | (0.000119) | (0.000120) |
|-------------|------------|------------|------------|------------|
| Education   | 0.108      | 0.671**    | 0.173      | 0.0688     |
|             | (0.254)    | (0.283)    | (0.345)    | (0.344)    |
| Job         | -0.266     | -0.618**   | -0.392     | -0.681**   |
|             | (0.218)    | (0.249)    | (0.330)    | (0.333)    |
| Car ownership | -0.0539  | 0.148      | 0.174      | 1.088**    |
|             | (0.268)    | (0.317)    | (0.416)    | (0.436)    |
| Population  | -0.534     | -0.754*    | -0.888*    | -0.425     |
|             | (0.347)    | (0.402)    | (0.509)    | (0.518)    |
| License     | 0.469      | 0.364      | 0.236      | 1.167*     |
|             | (0.488)    | (0.540)    | (0.654)    | (0.670)    |
| Alternative-specific constants ||||
| Constant    | 1.621      | 3.575**    | -3.488     | -5.185**   |
|             | (1.539)    | (1.752)    | (2.483)    | (2.514)    |
| Respondent  | 409        |            |            |            |
| Observations | 409       |            |            |            |
| Log-likelihood | -355.4373 |         |            |            |
| Pseudo R²   | 0.4375     |            |            |            |

*In the next section, we will apply the well-recognized kind of MNL to estimate the VOT.

It should be noted that in the RP survey of urban travel, unless the survey is conducted on a trip with a very similar attribute at the same time, it is difficult to estimate a reasonable time value. The approximate travel attributes here refer to the same departure and arrival places in a macro sense (eg, Guangzhou to Shenzhen), and similar departure times and costs are stable. This is relatively easy to sample for intercity and longer distance travel, but very difficult for highly mobile intra-city travel. Selecting a section of road is common practice in RP surveys, and one of the interstate highway HOT lanes in Minnesota is the site that has been used many times to estimate VOT through RP surveys, using loop detectors[70] or GPS instruments[71] Measure travel time data. Therefore, SP surveys with virtual scenarios with restricted travel attributes are more beneficial for us to collect data suitable for estimating the value of time.

### 5.2.2 VOT estimation in SP surveys

First, we give a set of utility functions with "bus" as the basic backup scheme, in which the utility functions of the basic backup scheme do not have ASC, and the utility functions of the other schemes have a fixed ASC, which does not change with the scheme. The purpose of this is to improve the significance of the estimated results. It has been proved many times that the ASC diversification of the utility function will make the estimated results unreasonable. The utility functions of this formal group will be used for the MNL model, MIXL model, S-MNL model, G-MNL model, LC model and MM-MNL model operation:

$$\begin{aligned} U^{Bus} &= \beta_1 Time + \beta_2 Cost + \varepsilon_{ni} \\ U^{Metro} &= \beta_1 Time + \beta_2 Cost + ASC + \varepsilon_{ni} \\ U^{Bike} &= \beta_1 Time + \beta_2 Cost + ASC + \varepsilon_{ni} \\ U^{Taxi} &= \beta_1 Time + \beta_2 Cost + ASC + \varepsilon_{ni} \\ U^{Drive\ Alone} &= \beta_1 Time + \beta_2 Cost + ASC + \varepsilon_{ni} \end{aligned} \quad (13)$$

We define the computational form of VTT, VTTS (WTP), and WTA based on the ideas put forward by Hess et al.:

$$VTT = \frac{\partial U / \partial Time_{SP}}{\partial U / \partial Cost_{SP}} = \frac{\beta_1}{\beta_2} \quad (14)$$

$$VTTS = WTP = \frac{\partial U / \partial Time_{SP\text{-}gain}}{\partial U / \partial Cost_{SP\text{-}loss}} = \frac{\beta_{1-gain}}{\beta_{2-loss}} \quad (15)$$

$$WTA = \frac{\partial U / \partial Time_{SP\text{-}loss}}{\partial U / \partial Cost_{SP\text{-}gain}} = \frac{\beta_{1-loss}}{\beta_{2-gian}} \quad (16)$$

$$BVTT = \sqrt{WTP \times WTA} = \sqrt{VTTS \times WTA} \quad (17)$$

where, alternative attribute items $Time$ and $Cost$ corresponds to $Time_{SP}$ and $Cost_{SP}$, $Time_{SP\text{-}gain}$ and $Cost_{SP\text{-}loss}$, $Time_{SP\text{-}loss}$ and $Cost_{SP\text{-}gain}$ in the estimation of VTT, VTTS (WTP) and WTA. $\beta$ is the parameter of utility function adapted to different models.

Table 5 shows the estimation of all data in the sample using all models including the LCL model and the MM-MNL model. As can be seen, the parameter estimates for the MNL model, MIXL model, S-MNL model and G-MNL model are all statistically significant and all appear negative. On the other hand, we noticed anomalies in the estimates from the LCL model and the MM-MNL model, with some terms showing as significant positives, insignificant negatives, or insignificant positives. The LCL model has five significant positive numbers (0.13588**/0.01179*/0.05819***/0.04463***/0.13042***), and the MM-MNL model has three significant positive numbers ( 0.07408***/0.05100**/0.14749***), these items have a positive effect on the traveler's utility, and the traveler expects a higher time difference or cost difference for the currently selected travel mode. Greene et al.[44] found that MNL and MIXL were more significant than LCL and MM-MNL when applying the Australian one-time shipment data. Specifically, certain attributes have statistical value only for one latent class, subject to the class probability. Further, we try to change the form of the utility function, change the constant item of the utility function, add some personal attribute items, etc., but it is difficult to correct this problem. At the same time, try to increase the number of categories to three, four or even more, and this anomaly will be exacerbated. Therefore, the results obtained by the two models using time and cost as the classification basis are not ideal and have certain limitations. The application of LPA/LCA just overcomes this shortcoming to some extent.

Table 5. Estimation result of all sample with DCMs

| Attribute | MNL Est. | SE | MIXL Est. | SE | S-MNL Est. | SE | G-MNL Est. | SE |
|---|---|---|---|---|---|---|---|---|
| Travel time (min) | -0.01803*** | -0.00395 | -0.07596*** | -0.02153 | -0.01805*** | 0.00399 | -0.07909*** | 0.02091 |
| Travel cost (CNY) | -0.0175** | -0.00712 | -0.21208*** | -0.05455 | -0.01747** | 0.00715 | -0.22973*** | 0.0565 |
| Gain of travel time (min) | -0.03845*** | -0.01048 | -0.20072*** | -0.0521 | -0.03973*** | 0.01202 | -0.26445*** | 0.05151 |
| Loss of travel cost (CNY) | -0.02077 | -0.02163 | -0.7486*** | -0.21848 | -0.0207 | 0.02246 | -0.88521*** | 0.18786 |
| Loss of travel time (min) | -0.03958*** | -0.01211 | -0.3351*** | -0.09408 | -0.0339*** | 0.01009 | -0.37631*** | 0.1234 |
| Gain of travel cost (CNY) | -0.01682 | -0.01411 | -0.53734*** | -0.14747 | -0.00678 | 0.0199 | -0.57776*** | 0.17537 |
| ASC | 0.40517*** | -0.15404 | 0.55645** | 0.3524 | 0.40704*** | 0.15517 | 0.54633*** | 0.20912 |
|  | 0.36405** | -0.15257 | 0.55474*** | -0.18182 | 0.37602** | 0.1651 | 0.48417*** | 0.17358 |
|  | 0.40656*** | -0.15218 | 0.35176 | -0.22465 | 0.4803*** | 0.17406 | 0.36895* | 0.21868 |
| $\tau$ |  |  |  |  | 0.12021 | 0.49923 | 0.94942*** | 0.23312 |
|  |  |  |  |  | 0.3702 | 0.56233 | 0.96230*** | 0.13761 |
|  |  |  |  |  | 0.40717 | 0.51386 | 0.69468* | 0.40814 |
| $\sigma$ |  |  |  |  | 0.99969*** | 0.12008 | 0.99639 | 1.19736 |
|  |  |  |  |  | 0.99864*** | 0.38072 | 0.99703 | 1.20347 |

|  |  |  | 0.99843** | 0.42162 | 0.99868 | 0.77954 |
|---|---|---|---|---|---|---|
|  |  |  |  |  | 0.72787*** | 0.25518 |
| γ |  |  |  |  | 0.49591*** | 0.16409 |
|  |  |  |  |  | 0.2951 | 0.38258 |

| | LCL | | | | MM-MNL | | | |
|---|---|---|---|---|---|---|---|---|
| | Class1 | | Class2 | | Class1 | | Class2 | |
| Attribute | Est. | SE | Est. | SE | Est. | SE | Est. | SE |
| Travel time (min) | -0.2989*** | 0.09083 | 0.01179* | 0.00613 | -0.06498*** | 0.01426 | -0.46883 | 3.10138 |
| Travel cost (CNY) | -1.16347*** | 0.40996 | 0.05819*** | 0.01309 | -0.67776*** | 0.25693 | 0.07408*** | 0.01435 |
| Gain of travel time (min) | 0.00765 | 0.0142 | -1.38889*** | 0.40064 | -0.03888 | 748.4215 | -0.03804 | 741.5637 |
| Loss of travel cost (CNY) | 0.13588*** | 0.03246 | -6.94397*** | 2.20164 | -0.02089 | 440.9166 | -0.02039 | 439.8502 |
| Loss of travel time (min) | -1.91709*** | 0.19468 | 0.04463*** | 0.01626 | -1.99291*** | 0.36152 | 0.05100*** | 0.01755 |
| Gain of travel cost (CNY) | -4.19503*** | 0.35417 | 0.13042*** | 0.02332 | -3.62254*** | 0.51299 | 0.14749*** | 0.02628 |
| ASC | | | | | 1.56575*** | 0.40073 | -7.87003 | 46.42047 |
| | | | | | 0.36415 | 6411.157 | 0.35686 | 6481.351 |
| | | | | | 1.72963 | 2.91339 | -0.38863* | 0.20084 |
| Classprobability | 0.32929*** | 0.04083 | 0.67071*** | 0.04083 | 0.57877*** | 0.05463 | 0.42123*** | 0.05463 |
| | 0.71125*** | 0.03553 | 0.28875*** | 0.03553 | 0.5 | 884637.3 | 0.5 | 884637.3 |
| | 0.31387*** | 0.03115 | 0.68613*** | 0.03115 | 0.31060*** | 0.03083 | 0.68940*** | 0.03083 |

After LPA, the data were classified into three categories, and all models except the LCL model and the MM-MNL model were used to estimate by category, and all the results are shown in Tables 6-8. LPA has replaced the classification function of LCL model and MM-MNL model, and the estimation results of these two models on small sample data are not good, so these two models are not used for calculation.

Here, it needs to be explained that the classification function of the MM-MNL model still cannot replace the LPA, mainly for the following reasons: MM-MNL requires panel data, and the actual traveler attribute data is long data, even if it is converted to wide type, because there is only one selection scenario , it cannot be calculated normally; after LPA, a classification label will be added to each group of data, such as the first group of data is classified as type 1, and the second group of data is classified as type 2, which obviously cannot be done if MM-MNL is used; MM - The MNL classification effect is not good, even if it degenerates into LCL, as the number of categories increases, the classification effect will drop off a cliff. After trying, after being divided into three categories, the parameters in each category are almost insignificant, and the category probability, AIC, BIC and other indicators are also very poor, which may be related to the sample size, but LPA still has a good effect on the three categories of data. Effect; further, whether MM-MNL can be applied well depends on the construction of the utility function. By constantly trying to change the structure of the utility function, even if there is a good effect in the end, it will be very troublesome and even deviate from the original intention of the classification.

From the results, the estimated parameters of the three categories of the four models are basically significant, but it is still found that the MNL model, the S-MNL model and the G-MNL model have Gain/Loss of travel in the estimation of type III. The two parameters of cost are positive numbers, and in the MNL model and G-MNL, the parameter of Gain of travel cost is greater than the parameter of Loss of travel cost. In addition, the parameter estimate of Travel cost by the MNL model is also positive, which reveals that the travel mode with higher travel cost does not reduce the choice tendency of the Type III population. On the contrary, the travel mode with higher travel cost brings convenience, Comfort, punctuality, etc. are more attractive. Hess[72] gives a corresponding explanation for the problem of positive time coefficients and positive cost coefficients when estimating VTTS using the MIXL model: it is often impossible to explicitly quantify the impact of joint activities or travel experience factors, so there is an extreme risk of biased estimates of travel time coefficients. Likely. At the same time, in the case that the model produces negative time/cost coefficients, the coefficients of the two can be shifted upwards or downwards from zero by adding some related attributes that are not included in the original utility function.

Table 6. Estimation result of type 1 with DCMs

| | MNL | | MIXL | | S-MNL | | G-MNL | |
|---|---|---|---|---|---|---|---|---|
| *Attribute* | Est. | SE | Est. | SE | Est. | SE | Est. | SE |
| Travel time (min) | -0.02684*** | -0.0081 | -0.13994 | -0.12408 | -0.05033** | 0.02233 | -0.09718* | 0.0562 |
| Travel cost (CNY) | -0.07519*** | -0.01848 | -0.83726 | -0.55053 | -0.17882*** | 0.03559 | -0.62078** | 0.29835 |
| Gain of travel time (min) | -0.06411*** | -0.02102 | -0.30648** | -0.11915 | -0.17581*** | 0.05132 | -0.31360* | 0.17552 |
| Loss of travel cost (CNY) | -0.24661*** | -0.06362 | -2.29796** | -0.94409 | -1.09943*** | 0.22544 | -2.29597** | 1.11777 |
| Loss of travel time (min) | -0.05257** | -0.02275 | -0.43415** | -0.18875 | -0.14694*** | 0.04334 | -0.56877*** | 0.09882 |
| Gain of travel cost (CNY) | -0.11908*** | -0.03904 | -1.65127** | -0.69577 | -0.97337*** | 0.24253 | -2.11210*** | 0.40889 |
|     | 0.51517* | -0.28581 | 0.3524  | -0.9397  | 0.6839    | 0.56489 | 0.61952   | 0.51539 |
| ASC | 0.52539* | -0.28415 | 0.53231 | -0.36616 | 3.12645*  | 1.60615 | 0.44597   | 0.52515 |
|     | 0.48186* | -0.28371 | 0.21803 | -0.42705 | 4.60044** | 2.1398  | 0.2281    | 0.31846 |
| τ   |         |          |         |          | 1.27745***| 0.16833 | 0.04095   | 0.62198 |
|     |         |          |         |          | 1.72799***| 0.16275 | 0.24857   | 0.90222 |
|     |         |          |         |          | 2.02409***| 0.19637 | 0.52356***| 0.17437 |
| σ   |         |          |         |          | 0.99628   | 2.02057 | 0.99929***| 0.04124 |
|     |         |          |         |          | 1.00537   | 3.85034 | 0.99932***| 0.25134 |
|     |         |          |         |          | 1.00937   | 5.68603 | 0.99804*  | 0.55453 |
| γ   |         |          |         |          |           |         | 23.2943   | 347.6285|
|     |         |          |         |          |           |         | 5.22734   | 14.0405 |
|     |         |          |         |          |           |         | -1.21262  | 0.90897 |

Table 7. Estimation result of type 2 with DCMs

|                           | MNL        |          | MIXL      |          | S-MNL      |         | G-MNL      |           |
| ------------------------- | ---------- | -------- | --------- | -------- | ---------- | ------- | ---------- | --------- |
| Attribute                 | Est.       | SE       | Est.      | SE       | Est.       | SE      | Est.       | SE        |
| Travel time (min)         | -0.02535***| -0.00675 | -0.11267  | -0.07249 | -0.02537***| 0.00676 | -0.09051   | 0.06726   |
| Travel cost (CNY)         | -0.02030*  | -0.01188 | -0.18169**| -0.09081 | -0.02033*  | 0.01188 | -0.17303*  | 0.10224   |
| Gain of travel time (min) | -0.05375***| -0.01799 | -0.37493**| -0.15994 | -0.05447***| 0.02099 | -0.37787*  | 0.19697   |
| Loss of travel cost (CNY) | -0.01613   | -0.03408 | -0.83877**| -0.39905 | -0.01589   | 0.03463 | -0.83002*  | 0.43542   |
| Loss of travel time (min) | -0.06456***| -0.02181 | -1.28610* | -0.66766 | -0.05627***| 0.01939 | -10.52622* | 0.83014   |
| Gain of travel cost (CNY) | -0.02186   | -0.0255  | -1.14249  | -0.83524 | -0.00992   | 0.03279 | -10.34551* | 0.78731   |
|                           | 0.66530**  | -0.26717 | 0.76889   | -0.55879 | 0.66635**  | 0.26789 | 1.09733    | 0.76735   |
| ASC                       | 0.60771**  | -0.26369 | 0.79399** | -0.36249 | 0.62090**  | 0.30351 | 0.97753**  | 0.40489   |
|                           | 0.65972**  | -0.26387 | 0.24035   | -0.43543 | 0.84210**  | 0.33183 | 0.22599    | 0.60173   |
| τ                         |            |          |           |          | 0.04396    | 0.53297 | 0.00029    | 1.45759   |
|                           |            |          |           |          | 0.22266    | 1.25817 | 0.00011    | 1.47395   |
|                           |            |          |           |          | 0.41037    | 0.5162  | 0.33317    | 0.60078   |
| σ                         |            |          |           |          | 0.99974*** | 0.04379 | 1.00000*** | 0.00029   |
|                           |            |          |           |          | 0.99858*** | 0.22498 | 1.00000*** | 0.00011   |
|                           |            |          |           |          | 0.99733**  | 0.42827 | .99939***  | 0.34113   |
| γ                         |            |          |           |          |            |         | -1074.49   | 0.2591D+08|
|                           |            |          |           |          |            |         | 1852.13    | 0.2591D+08|
|                           |            |          |           |          |            |         | 1.56413    | 20.31903  |

Table 8. Estimation result of type 3 with DCMs

|                           | MNL        |          | MIXL     |          | S-MNL      |         | G-MNL       |         |
| ------------------------- | ---------- | -------- | -------- | -------- | ---------- | ------- | ----------- | ------- |
| Attribute                 | Est.       | SE       | Est.     | SE       | Est.       | SE      | Est.        | SE      |
| Travel time (min)         | -0.01421** | -0.00615 | -0.14471 | -0.10268 | -0.01421** | 0.00622 | -0.05047*   | 0.02638 |
| Travel cost (CNY)         | 0.01877*   | -0.00972 | -0.05223 | -0.05813 | 0.01877**  | 0.00894 | -0.00432    | 0.02597 |
| Gain of travel time (min) | -0.02768*  | -0.01603 | -0.36445 | -0.30182 | -0.02768*  | 0.0159  | -0.23873*** | 0.08822 |
| Loss of travel cost (CNY) | 0.09779*** | -0.03057 | -0.1424  | -0.28452 | 0.09779*** | 0.02953 | 0.06265     | 0.12517 |
| Loss of travel time (min) | -0.03150* | -0.01793 | -0.27613** | -0.12439 | -0.01314 | 0.01509 | -0.34104* | 0.19151 |
| Gain of travel cost (CNY) | 0.03620* | -0.01858 | -0.01139 | -0.07596 | 0.12010*** | 0.02852 | 0.00764 | 0.11536 |
| | 0.67408** | -0.27526 | 0.16181 | -0.5882 | 0.67408** | 0.2747 | 0.75787** | 0.33448 |
| ASC | 0.65424** | -0.27315 | 0.35754 | -0.50187 | 0.65424** | 0.2706 | 0.62399* | 0.32543 |
| | 0.79337*** | -0.27052 | 0.60904* | -0.34363 | 0.73380*** | 0.27278 | 0.62426* | 0.36345 |
| | | | | | 0 | 1.31884 | 0 | 1.34797 |
| $\tau$ | | | | | 0 | 1.13923 | 0.74552** | 0.34521 |
| | | | | | 0 | 1.13957 | 1.12081*** | 0.30996 |
| | | | | | 1.00000*** | 0.3495D-07 | 1.00000*** | 0.2581D-07 |
| $\sigma$ | | | | | 1.00000*** | 0.1825D-07 | 0.99745 | 0.84643 |
| | | | | | 1.00000*** | 0.3161D-07 | 0.9927 | 1.49126 |
| | | | | | | | 0.98309 | 0.7792D+12 |
| $\gamma$ | | | | | | | -2.12955 | 2.31852 |
| | | | | | | | 0.11566 | 0.317 |

Finally, we summarize the results summarized above. First, the time and cost of travel are perceived differently by people from different classes. The "student group" expects less travel costs, while the "social backbone group" expects savings in travel time; second, Gain/Loss of travel time and Gain/Loss of travel cost give the absolute value of the traveler's willingness to travel for the selected reference travel mode in terms of travel time and cost. Less satisfied with time/travel cost. Of course, it is difficult to find a faster and cheaper way to travel in reality. Travel time and travel cost balance each other. The magnitude and positive or negative of the parameters of the difference item reflect travelers' attitudes toward travel time/travel cost. Secondly, $\tau$, $\sigma$ and $\gamma$ are the three parameters of the G-MNL model to calibrate the MNL model, MIXL model and S-MNL model. Since we do not have them to constrain, it appears that the estimation result of one model is similar to the estimation result of another model. , which is further explained in the study by Fiebig[41]. The estimation results of the MNL model and the S-MNL model in Type II and III confirm this well.

Table 9. Summary of VOT

| VOT | MNL | MIXL | S-MNL | G-MNL | LCL | | MM-MNL | |
|---|---|---|---|---|---|---|---|---|
| | | | | | Class 1 | Class 2 | Class 1 | Class 2 |
| *Total(Average income is 28.90 CNY/h)* | | | | | | | | |
| VTT (CNY/h) | 61.82 | 21.49 | 61.99 | 20.66 | 15.41 | 12.16 | 5.75 | -379.72 |
| VTTS/WTP (CNY/h) | 111.07 | 16.09 | 115.16 | 17.92 | **3.38** | 12.00 | 111.67 | 111.94 |
| WTA (CNY/h) | 141.19 | 37.42 | 300.00 | 39.08 | 27.42 | 20.53 | 33.01 | **20.75** |
| BVTT (CNY/h) | 125.23 | 24.53 | 185.87 | 26.47 | **9.62** | 15.70 | 60.71 | **48.19** |
| *Type 1 (Average income is 6.41 CNY/h)* | | | | | | | | |
| VTT (CNY/h) | 21.42 | 10.03 | 16.89 | 9.39 | | | | |
| VTTS/WTP (CNY/h) | 15.60 | 8.00 | 9.59 | 8.20 | | | | |
| WTA (CNY/h) | 26.49 | 15.78 | 9.06 | 16.16 | | | | |
| BVTT (CNY/h) | 20.33 | 11.24 | 9.32 | 11.51 | | | | |
| *Type 2 (Average income is 33.78 CNY/h)* | | | | | | | | |
| VTT (CNY/h) | 74.93 | 37.21 | 74.87 | 31.39 | | | | |
| VTTS/WTP (CNY/h) | 199.94 | 26.82 | 205.68 | 27.32 | | | | |
| WTA (CNY/h) | 177.20 | 67.54 | 340.34 | 61.05 | | | | |
| BVTT (CNY/h) | 188.23 | 42.56 | 264.58 | 40.84 | | | | |
| *Type 3 (Average income is 46.20 CNY/h)* | | | | | | | | |
| VTT (CNY/h) | -45.42 | 166.24 | -45.42 | 700.97 | | | | |
| VTTS/WTP (CNY/h) | -16.98 | 153.56 | -16.98 | -228.63 | | | | |
| WTA (CNY/h) | -52.21 | 1454.59 | -6.56 | -2678.32 | | | | |
| BVTT (CNY/h) | **29.78** | 472.62 | **10.56** | 782.53 | | | | |

*Bold estimates : WTA and WTP are not both negative

Based on the estimated values of all the models given in Tables 5-8 for the three groups of people, we have statistically summarized the time values of this study, as shown in Table 9.

Among the 20 sets of VOT results of the six models, a total of 15 sets of results show that the absolute value WTA>WTP (or VTTS), which is a good demonstration of the perspective of prospect theory, indicating that most travelers need more tickets in the face of increased travel time. price compensation. In addition, the size of the time value and the type of the group have a high degree of adaptation. The time value generally includes "social backbone group" > "elderly group" > "student group", in particular, there are three time values for "social backbone group" In addition to the MIXL model, there are multiple results showing negative numbers, which are difficult to interpret. The estimation results of different models of the same type are very different. By introducing the hourly wages of the three groups as the comparison basis, it is found that in addition to the estimation of the WTA of type III, the MIXL model is more versatile and the estimated results are more reasonable. . At the same time, note that the four models have good estimation results for the three time values of type I. This is similar to the evidence for VTTS estimated from large-scale household surveys in Japan[73], who concluded that the VTTS of the youth group and the elderly group is lower than that of other age groups. Schmid[74] also reported VTTS results in relation to traveler attributes, with higher VTTS (4.6 EUR/hour) in the high-income group than in the less educated group (4.3 EUR/hour). But obviously, our explanation of VOT by using LPA comprehensively considering the attributes of the respondents will be more convincing.

Due to the asymmetry of WTP and WTA, the estimates of VTT and VTTS by the six models are all very significantly different, and Table 10 presents the degree of difference between VTT and VTTS. From the table, it is not difficult to see that, in contrast, there is a smaller difference between the VTT and VTTS estimated by the MIXL model and the G-MNL model, while the VTTS of the MM-MNL for Class 1 is more than 18 times that of the VTT. In addition, the overall differences in VTT and VTTS of type I were smaller than those of the other three categories. This can further illustrate that, in addition to responding to the difference between VTT and VTTS, the choice of model and sample can have a large impact on the difference between the two estimates. There is a big gap between our calculation results and the WTA and WTP ratios of Ramjerdi[31] and Daly & Hess[34] at 1.0~2.0 times. Daly & Hess compared work and non-work travel purposes and multiple travel modes and found that the difference between VTTS and VTT ranged from 0 to 110.70%. Therefore, if the difference between the two is not correctly distinguished, the application to the formulation of transportation policy may cause a large gap between the actual economic assessment and the economic assessment in the document, thereby increasing the risk of economic loss.

Table 10. Summary of bias between VTT & VTTS

|  | MNL | MIXL | S-MNL | G-MNL | LC | | MM-MNL | |
|---|---|---|---|---|---|---|---|---|
|  |  |  |  |  | Class 1 | Class 2 | Class 1 | Class 2 |
| *Total* | | | | | | | | |
| Bias between VTT & VTTS | 79.68% | -25.14% | 85.77% | -13.23% | -78.09% | -1.28% | 1841.26% | -129.48% |
| Bias between BVTT & VTTS | -11.30% | -34.43% | -38.04% | -32.27% | -64.90% | -23.55% | 83.93% | 132.28% |
| *Type 1* | | | | | | | | |
| Bias between VTT & VTTS | -27.17% | -20.20% | -43.18% | -12.75% |  |  |  |  |
| Bias between BVTT & VTTS | -23.26% | -28.78% | 2.92% | -28.78% |  |  |  |  |

|  |  |  |  |  |
|---|---|---|---|---|
| *Type 2* | | | | |
| Bias between VTT & VTTS | 166.85% | -27.92% | 174.69% | -12.97% |
| Bias between BVTT & VTTS | 6.22% | -36.99% | -22.26% | -33.11% |
| *Type 3* | | | | |
| Bias between VTT & VTTS | -62.61% | -7.63% | -62.61% | -132.62% |
| Bias between BVTT & VTTS | -157.03% | -67.51% | -260.85% | -129.22% |

# 6 Conclusion

This paper evaluates RP-SP data from Guangzhou residents' intra-city travel preferences using various Logit models. Respondents were given a choice between three virtual scenarios and in each scenario, transit, subway, bicycle (electric vehicle), taxi, and self-driving car, and two tests of willingness were conducted simultaneously. Then, we applied the LPA model to divide the respondents into three groups: "student group", "social backbone group" and "elderly group", and estimated the VTT, VTTS (WTP) and WTA of the three types of groups, respectively. The time value results of this paper can provide more new insights for transportation policy makers and managers in the formulation of transportation investment and public transportation fares. More importantly, we clarify the different definitions of VTT and VTTS and different estimation methods, reducing unnecessary controversy over these two terms.

(1) We use prospect theory to define VTT and VTTS and calculate VOT, obtain VTT, WTP (VTTS) and WTA, and compare the differences of VTT and VTTS under different types of groups and different models. At the same time, this paper also emphasizes that different models and different data samples will have large gaps in calculating VOT.

(2) The time value of the three types of groups "student group", "social backbone group" and "senior group" obtained by LPA are quite different, such as "social backbone group" > "elderly group" > "student group". Cheaper transportation is more attractive to the "student group" and "elderly group", while the "societal backbone group" is more inclined to choose a faster way of travel. Therefore, vigorously developing public transportation can benefit more social groups. At the same time, considering that high-income groups in society have higher time value, it is also necessary to provide them with more comfortable and fast travel methods to reduce their commuting losses.

(3) The results reflect that the row group is generally more averse to time loss, that is, VTTS < WTA. Therefore, it seems more welfare to reduce unnecessary time lost by travelers than to try to use larger investment to promote the speed of travel. Specifically, it is more beneficial to the overall travel of the society to alleviate the problem of congestion during the rush hour through some effective traffic measures.

(4) In previous studies, it is usually the reference between one mode of travel and another mode of travel. In the case of binary choice and multi-choice, the method in this paper is difficult to implement, because it is difficult to know the reference point in the traveler's psychology. Which travel mode, this will be very troublesome to estimate VTT and VTTS. It is worth thinking about whether the reference point between the "expectation" reference point of this paper and the mode of travel is better or worse, which remains to be explored.

(5) Finally, the data in this study comes from residents of Guangzhou City. Due to the influence of the economic, traffic, social and environmental conditions in the region, the travel preferences of residents in developed cities may be different from those in remote areas and rural areas. As a result, the VTT, VTTS (WTP) and WTA results of residents in the two places were different. Therefore, in order to fully prove the conclusion of this paper, more domestic cities and regions should be studied.